RESEARCH ARTICLE

## Molecules as Sources for Indistinguishable Single Photons

Ville Ahtee[ab], Robert Lettow[a], Robert Pfab[a], Alois Renn[a], Erkki Ikonen[b], Stephan Götzinger[a*], and Vahid Sandoghdar[a]

[a] *Laboratory of Physical Chemistry and optETH, ETH Zürich, CH-8093 Zürich, Switzerland*
[b] *Metrology Research Institute, Helsinki University of Technology (TKK) and Centre for Metrology and Accreditation (MIKES), P.O. Box 3000, FI-02015 TKK, Finland*

We report on the triggered generation of indistinguishable photons by solid-state single-photon sources in two separate cryogenic laser scanning microscopes. Organic fluorescent molecules were used as emitters and investigated by means of high resolution laser spectroscopy. Continuous-wave photon correlation measurements on individual molecules proved the isolation of single quantum systems. By using frequency selective pulsed excitation of the molecule and efficient spectral filtering of its emission, we produced triggered Fourier-limited single photons. In a further step, local electric fields were applied to match the emission wavelengths of two different molecules via Stark effect. Identical single photons are indispensible for the realization of various quantum information processing schemes proposed. The solid-state approach presented here prepares the way towards the integration of multiple bright sources of single photons on a single chip.

---

*Corresponding author. Email: stephan.goetzinger@phys.chem.ethz.ch

# 1. Introduction

Interference effects of light produced by two independent, unrelated sources were already studied in 1955 [1]. Since then experiments have become more sophisticated and detectors more sensitive, showing that interference exists even for a few photons. In fact, it was as late as 1983 that Mandel theoretically investigated interference and correlation effects of single photons emitted by independent sources [2], likely inspired by new experimental techniques for studying resonance fluorescence from single atoms. First experiments in this direction utilized a spontaneous parametric down-conversion (SPDC) process where indistinguishable photon pairs are generated from higher frequency pump photons in a nonlinear crystal [3]. However, such photons cannot be considered fully independent since they were produced by the same pump photons and the same region of a crystal in a coherent process. In a step towards independent sources, photons were created in spatially separated crystals [4], but since the crystals were still pumped by the same laser, the resulting photons are still not completely independent. Only very recently indistinguishable photons from independent SPDC sources were reported where different crystals were pumped by electrically synchronized but otherwise independent femtosecond lasers [5]. An inherent drawback of SPDC sources remains, however, that their photon statistics is governed by a Poissonian distribution. In other words, the probability of emitting more than one photon can only be reduced, but it is never zero. As a result, one is limited to a weak photon flux to ensure the creation of single photon pairs. Furthermore, photons created by SPDC are spectrally broad and additional spectral filtering is often needed, resulting in a further reduction of the photon flux.

Motivated by the limitations of SPDC, active research has been undertaken to develop true single-photon sources that are capable of emitting indistinguishable photons. Photon indistinguishability is indispensable in any realization of quantum computation with single photons [6]. It is important to note that not only consecutively emitted photons from one emitter need to be identical, but also photons from different sources must be indistinguishable. In fact, schemes proposing efficient quantum computation require multiple single-photon sources emitting mutually indistinguishable single photons. Several two-level emitters, such as single quantum dots [7], single atoms in cavities [8], and single molecules [9] have successfully been operated as sources for indistinguishable single photons. In these experiments, consecutively emitted photons of the very same emitter where overlapped on a beam splitter. In another approach, atomic ensembles have been used for the generation of indistinguishable photon pairs [10]. Very recently a milestone was reached when indistinguishable photons were produced from independently trapped atomic emitters [11, 12]. Atoms and ions are very attractive for demonstrating proofs of principle. However, trapping single atoms requires sophisticated technology and the experimental procedure becomes particularly challenging if a large number of independently trapped atoms is desired.

Emitters in solid matrices offer a promising alternative. The position of the emitter is fixed so that no elaborate trapping techniques are required. The resulting devices can be very small and densely packed due to the monolithic nature of a solid-state system, and electrical pumping is often possible [13]. Semiconductor quantum dots, for example, are attractive for several reasons. First, they are well known to be photostable. Second, fabrication technology from semiconductor industry is readily available for micro- and nano-processing of chips that contain hundreds of quantum dots. On the other hand, their photon indistinguishablity is limited by dephasing [14] and their linewidth is usually not lifetime limited. As a solution, it is proposed to switch off spontaneous emission and use schemes similar to a stimulated Raman adiabatic passage for generating photons [14, 15]. The experimental realization can be challenging and the proof of principle has not yet been demonstrated.

Organic molecules in a crystalline host matrix offer another powerful system. At low temperatures some molecular transitions become lifetime limited and offer almost unity quantum yield. Molecules, like any other system that generates photons by spontaneous emission, offer another important advantage compared to SPDC. Even though the same laser excites two molecules, coherence is destroyed by the spontaneous emission process, making the emitted photons truly independent. Recently we realized two independent Fourier-limited solid-state single-photon sources [16]. The solid-state arrangement of this approach enables nearly indefinite measurement times using the same emitter and a straightforward method for scaling and miniaturization. This paper characterizes the system further, reports on our recent developments, and discusses future prospective.

**2. Method**

Figure 1(a) shows a schematic of the experimental setup. Two independent low temperature microscopes in a liquid helium bath cryostat ($T = 1.4K$) are separated by an opaque wall. As excitation sources we used either a continuous wave (cw) or a pulsed dye laser. Both lasers could be coarsely frequency tuned by several tens of nanometers around 590 nm. The cw laser system (Coherent 899-29) is continuously tunable over several hundred GHz and has a linewidth of about 1 MHz. The pulsed laser produces pulses of 700 ps with a repetition rate of 76 MHz. A cavity dumper can reduce this rate by a factor of 20 to guarantee well separated photons emitted by a single molecule. Light from both lasers is coupled into a polarization maintaining single mode fiber and directed to the microscope objectives via galvo-optic mirror scanners and telecentric lens systems to provide beam scanning confocal microscopes. As emitters we used dibenzanthanthrene (DBATT) molecules embedded in Shpolskii matrices of n-tetradecane with a concentration of about $10^{-7}$ mol/l. The samples were sandwiched between hemispherical cubic zirconia solid immersion lenses (SIL) and glass substrates that contained interdigitating gold electrodes for Stark shifting the resonances of the molecules. Voltages up to 90 V, equivalent to electric fields of up to $5 \times 10^6$ V/m, could be applied to the electrodes spaced by 18 μm.

Aspheric lenses with a numerical aperture (NA) of 0.55 were placed inside the cryostat to focus the excitation beams onto the SILs and the samples [16, 17]. Such a lens combination leads to an overall NA of 1.12. The sample/SIL package was mounted on a home made three-axis piezo electric slider system that enabled adjustments with sub-micrometer precision. The emission from the molecules was collected by the same SIL/aspheric lens combination and was directed to an avalanche photodiode (APD), a spectrometer, an imaging CCD camera, or a Hanbury Brown and Twiss (HBT) type photon correlator. A set of optical interference filters cut out the excitation wavelength and other unwanted spectral components. Figure 1(b) shows a photograph of the sample holder. The diameter of the holder is 50 mm limited by the dimensions of the cryostat chamber.

There are two principal ways to excite the narrow band molecules at cryogenic temperatures. The first method excites the ground state molecule into the $S_{1,v=0}$ state (see Figure 2(a)). From there the molecule decays into the ground state $S_{0,v=0}$, and into a manifold of states with $v \neq 0$. These red shifted photons are detected, but the photons from the narrow 0-0 zero phonon line (ZPL) cannot be separated from the excitation laser. This technique of photoluminescence excitation spectroscopy can directly measure the linewidth of the 0-0 ZPL by scanning a narrowband laser across its resonance [18]. In a second method, one excites the molecule into a higher vibrational level of the electronically excited state ($S_{1,v=1}$). This state decays nonradiatively within a few ps into $S_{1,v=0}$, from where the molecule decays, giving access to photons of the lifetime limited 0-0 ZPL. Associated with the short lifetime of $S_{1,v=1}$ is a broadening of the energy level to about 30 GHz. Therefore, we need about two orders of magnitude larger pump power (100 nW) compared to conventional excitation of molecules via the 0-0 ZPL. Vibronic excitation also severely decreases the spectral selectivity. Molecules with 0-0 ZPLs separated by more than 10 GHz can be simultaneously excited owing to the broad $S_{1,v=1}$ level. For this reason we dilute our samples by about two orders of magnitude compared to commonly used samples in low temperature single molecule spectroscopy. Figure 2(b) shows a typical emission spectrum of a single DBATT molecule in n-tetradecane under vibronic excitation. The spectrum is dominated by the narrow 0-0 ZPL. In fact, we observe about 30% of the emitted photons in the 0-0 ZPL [16].

## 3. System characterization

Figure 3(a) shows an image recorded by one of the confocal microscopes over an area of 10 µm x 10 µm. Molecules are excited on 0-1 transitions and their red shifted fluorescence is collected. Several single molecules can be distinguished. Relatively high background fluorescence caused by the increase in the laser power reveals the shape of the crystalline host matrix with characteristic cracks. Figure 3(b) displays the high spatial resolution of the system. It shows a 2.5 µm x 2.5 µm scan around the molecule marked in Figure 3(a). A Gaussian profile is fitted to the cross section of the graph. The resulting full width at half-maximum (FWHM) of 330 nm indicates that the diameter of the laser focus is close to the theoretical diffraction-limited value of 270 nm expected from a NA=1.12 objective. Such a tight focus is rather important for suppressing the probability of exciting more than one molecule via the broad vibronic excited state.

After having identified a molecule that is spatially well separated from others (see figure 3(b)), we confirmed the lifetime limited linewidth of the 0-0 ZPL by photoluminescence excitation spectroscopy. Typically we measured 18 MHz full width at half-maximum, a value that corresponds well to the excited state lifetime of 9.4±1 ns reported in literature [19]. Photon correlation measurements using the HBT setup proved the isolation of a single quantum system. Figure 4(a) shows the result of such a measurement performed on a single molecule excited by the cw laser. Antibunching is clearly visible with $g^2(0) = 0.34$. The antibunching decay time of 8.1 ns is a bit shorter than the lifetime of the molecule, indicating that the system was not operated in the weak excitation limit.

In order to realize a triggered single-photon source, we excited the same molecule with the pulsed laser (Figure 4(b)). As in the case of the cw measurement, we excited the molecule via $S_{1,v=1}$ and detected its fluorescence on the 0-0 ZPL. Again we observe antibunching with $g^2(0)<0.5$. The two-photon emission probability is reduced to 44% compared to a Poissonian source. The remaining coincidence counts at t=0 can possibly be attributed to a weak excitation of other slightly detuned molecules in the neighbourhood. This performance could be improved by diluting the sample further and incorporating a pinhole in the detection path to ensure exclusive collection of photons from the molecule of interest. To our knowledge, these results constitute the first report of single molecule spectroscopy where pulsed excitation is combined with spectral and spatial resolution. The technical difficulty in this detection scheme is to obtain pulses that are shorter than the fluorescence lifetime but long enough to avoid the excitation of several molecules at the same time. We achieved this by using a linear dye laser pumped by a frequency-doubled Nd:YVO$_4$ laser with a pulse width of about 13 ps. We then stretched the pulses to about 700 ps by implementing two etalons in the dye laser cavity.

Having shown that the selected molecule has a Fourier-limited linewidth and can be used as a triggered single-photon source, we looked for another molecule in the second microscope. After we found a promising candidate with a 0-0 ZPL within a few GHz of the first molecule's emission line, we varied the voltage applied to the Stark electrodes in one of the microscopes. Figure 5(a) shows an example of Stark tuning where 0-0 ZPL fluorescence excitation spectra were taken simultaneously from both samples as a function of the applied voltage. At zero volts, the 0-0 ZPLs of the molecules are separated by 180 MHz. With increasing voltage the 0-0 ZPL of one of the molecules shifts towards lower frequencies. At a voltage of 42 V, the resonances of the two molecules can no longer be distinguished.

True quantum indistinguishability requires identical photons with respect to wavelength, polarization state, and spatial mode. DBATT molecules embedded in a Shpolskii matrix have well-oriented transition dipole moments whose emission can be linearly polarized [17]. Thus it should not be a difficult task to obtain the same polarization state for both molecules. Indeed, we could successfully couple the 0-0 ZPL emission from single molecules into a polarisation maintaining single mode fibre with an efficiency of 30%, corresponding to more than $10^5$ photons per second.

Our current experiments involve only two molecules in different setups. However, one can easily imagine realizing a dense configuration of electrodes that are individually addressable on one chip as illustrated in Figure 5(b). Small solid immersion lenses together with microlens arrays would ensure a high collection efficiency of photons emitted by a large number of single-photon sources. Coupling the photons in a bundle of single mode fibres with polarization control and variable delay lines, would enable a device capable of simultaneously generating many indistinguishable single photons on demand.

## 4. Conclusion

In this paper we presented the realization of two identical Fourier-limited solid-state single-photon sources. Organic dye molecules were identified by cryogenic high resolution optical microscopy in two independent setups. The utilization of solid immersion lenses together with aspheric lenses was essential to achieve a tight excitation focus and thus a high degree of spatial selectivity. This arrangement also provides large collection efficiency essential for bright single-photon sources. Triggered single-photon emission was realized by introducing a frequency selective pulsed vibronic excitation scheme. The Stark effect was exploited to shift the transition frequency of a given molecule and thus achieve two independent single-photon sources with perfect spectral overlap. The high photostability and the compact solid-state arrangement of the sources allow nearly indefinite measurement times on the same emitter and a straightforward up-scaling to a larger number of identical single-photon sources. These results are promising for miniaturization and integration of several single-photon sources to even smaller volumes on a single chip. Further control and enhancement of the radiation properties of such single-photon sources can be envisioned by using microcavities [20] or nano-antennae [21, 22].

**Acknowledgement**

This work was supported by the ETH Zurich via the INIT program Quantum Systems for Information Technology (QSIT), the Swiss National Science Foundation (SNF) and the Academy of Finland. We thank G. Wrigge and I. Gerhardt for experimental help.

**References**

[1] Forrester, A.T.; Gudmundsen, R. A.; Johnson, P. O. Photoelectric Mixing of Incoherent Light. *Phys. Rev.* **1955**, 99,1691-1700.

[2] Mandel, L. Photon Interference and Correlation Effects Produced by Independent Quantum Sources. *Phys. Rev. A* **1983,** 28, 929-943.

[3] Hong, C.K.; Ou, Z.Y.; Mandel, L. Measurement of Subpicosecond Time Intervals between Two Photons by Interference. *Phys. Rev. Lett.* **1987,** 59, 2044-2046.

[4] de Riedmatten, H.; Marcikic, I.; Tittel, W.; Zbinden, H.; Gisin, N. Quantum Interference with Photon Pairs Created in Spatially Separated Sources. *Phys. Rev. A* **2003,** 67, 022301.

[5] Kaltenbaek, R.; Blauensteiner, B.; Żukowski, M.; Aspelmeyer, M.; Zeilinger, A. Experimental Interference of Independent Photons. *Phys. Rev. Lett.* **2006,** 96, 240502.

[6] Knill, E.; Laflamme, R.; Milburn, G.J. A Scheme for Efficient Quantum Computation with Linear Optics. *Nature* **2001,** 409, 46-52.

[7] Santori, C.; Fattal, D.; Vučković, J.; Solomon, G.S.; Yamamoto, Y. Indistinguishable Photons from a Single-Photon Device. *Nature* **2002,** 419, 594-597.

[8] Legero, T.; Wilk, T.; Hennrich, M.; Rempe, G.; Kuhn, A. Quantum Beat of Two Single Photons. *Phys Rev. Lett.* **2004,** 93, 070503.

[9] Kiraz, A.; Ehrl, M.; Hellerer, Th.; Müstecaplıoğlu, Ö.E., Bräuchle, C.; Zumbusch, A. Indistinguishable Photons from a Single Molecule. *Phys. Rev. Lett.* **2005,** 94, 223602.

[10] Thompson, J.K.; Simon, J.; Loh, H.; Vuletić, V. A High-Brightness Source of Narrowband, Identical-Photon Pairs. *Science* **2006,** 313, 74-77.

[11] Beugnon, J.; Jones, M.P.A.; Dingjan, J.; Darquié, B.; Messin, G.; Browaeys, A.; Grangier, P. Quantum Interference between Two Single Photons Emitted by Independently Trapped Atoms. *Nature* **2006,** 440, 779-782.

[12] Maunz, P.; Moehring, D.L.; Olmschenk, S.; Younge, K.C.; Matsukevich, D.N.; Monroe, C. Quantum Interference of Photon Pairs from Two Remote Trapped Atomic Ions. *Nature Phys.* **2007,** 3, 538-541.

[13] Yuan, Z.; Kardynal, B. E.; Stevenson, R. M.; Shields, A. J.; Lobo, C. J.; Cooper, K.;. Beattie, N. S.; Ritchie, D. A.; Pepper, M. Electrically Driven Single-Photon Source. *Science* **2002**, 295, 102–105.

[14] Kiraz, A.; Atatüre, M.; Imamoğlu, A. Quantum-Dot Single-Photon Sources: Prospects for Applications in Linear Optics Quantum-Information Processing. *Phys. Rev. A* **2004**, 69, 032305.

[15] Yao, W.; Liu, R.-B.; Sham, L. J. Theory of Control of the Spin-Photon Interface for Quantum Networks. *Phys. Rev. Lett.* **2005**, 95, 030504.

[16] Lettow, R.; Ahtee, V.; Pfab, R.; Renn, A.; Ikonen, E.; Götzinger, S.; Sandoghdar, V. Realization of Two Fourier-Limited Solid-State Single-Photon Sources. *Opt. Express* **2007,** 15, 15842-15847.

[17] Wrigge, G.; Gerhardt, I.; Hwang, J.; Zumhofen, G.; Sandoghdar, V. Efficient Coupling of Photons to a Single Molecule and Observation of its Resonance Fluorescence. *Nature Phys.* **2008,** 4, 60 - 66.

[18] Orrit, M.; Bernard, J. Single Pentacene Molecules Detected by Fluorescence Excitation in a *p*-Terphenyl Crystal. *Phys. Rev. Lett.* **1990**, 65, 2716-2719.

[19] Boiron, A.-M.; Lounis, B.; Orrit, M. Single Molecules of Dibenzanthanthrene in *n*-hexadecane. *J. Chem. Phys.* **1996,** 105, 3969-3974.

[20] Press, D.; Götzinger, S.; Reitzenstein, S.; Hofmann, C.; Löffler, A.; Kamp, M.; Forchel, A.; and Yamamoto, Y. Photon Antibunching from a Single Quantum-Dot-Microcavity System in the Strong Coupling Regime. *Phys. Rev. Lett.* **2007**, 98, 117402.

[21] Kühn, S.; Håkanson, U.; Rogobete, L.; and Sandoghdar, V. Enhancement of Single-Molecule Fluorescence Using a Gold Nanoparticle as an Optical Nanoantenna. *Phys. Rev. Lett.* **2006**, 97, 017402.

[22] Rogobete, L.; Kaminski, F.; Agio, M.; Sandoghdar, V. Design of Plasmonic Nanoantennae for Enhancing Spontaneous Emission. *Opt. Lett.* **2007**, 32, 1623-1625.

# List of Figures

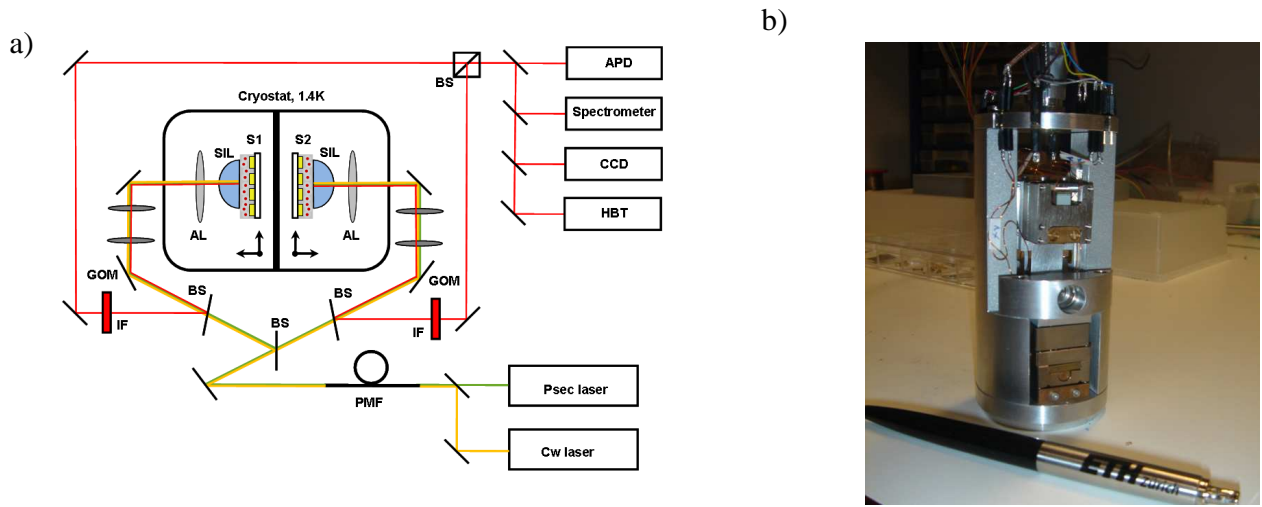

Figure 1. (a) Experimental setup: Two low temperature microscopes with solid immersion lenses (SIL) are placed in a liquid helium bath cryostat. Light from the excitation lasers is coupled into a polarization maintaining fiber (PMF) and guided to the microscopes equipped with galvo-optic mirror scanners (GOM) and telecentric lens systems. Fluorescence is collected by the same optical arrangement and directed to different detectors. S: sample, AL: aspheric lens, BS: beam splitter, IF: set of interference filter, APD: Avalanche photo detector, HBT: Hanbury Brown and Twiss autocorrelator. Further details are given in the text. (b) Photograph of the low temperature sample holder with piezo-electric slider positioners and AL-SIL imaging system.

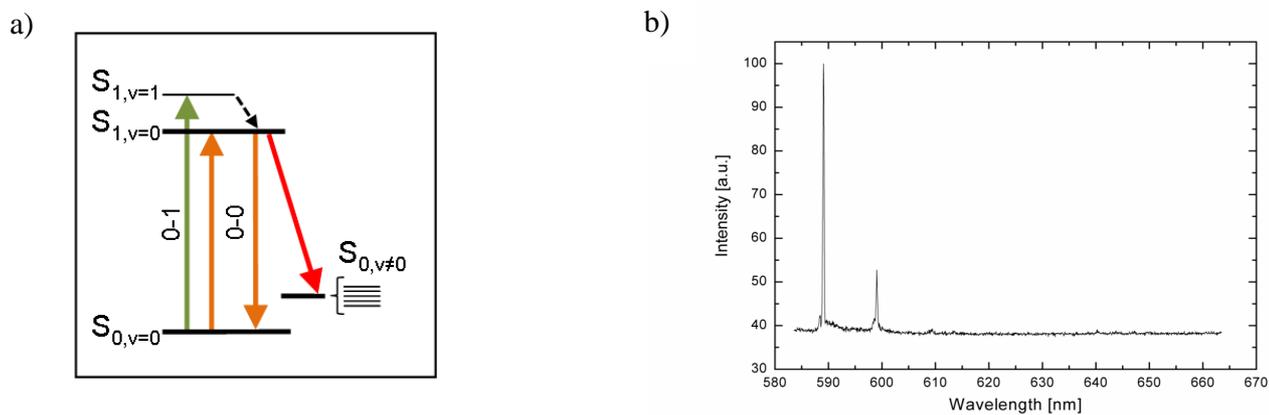

Figure 2. (a) Jablonsky diagram of a dye molecule with relevant energy levels. (b) Typical emission spectrum of a single DBATT molecule in n-tetradecane under vibronic excitation. The spectrum is dominated by the 0-0 ZPL at about 590 nm.

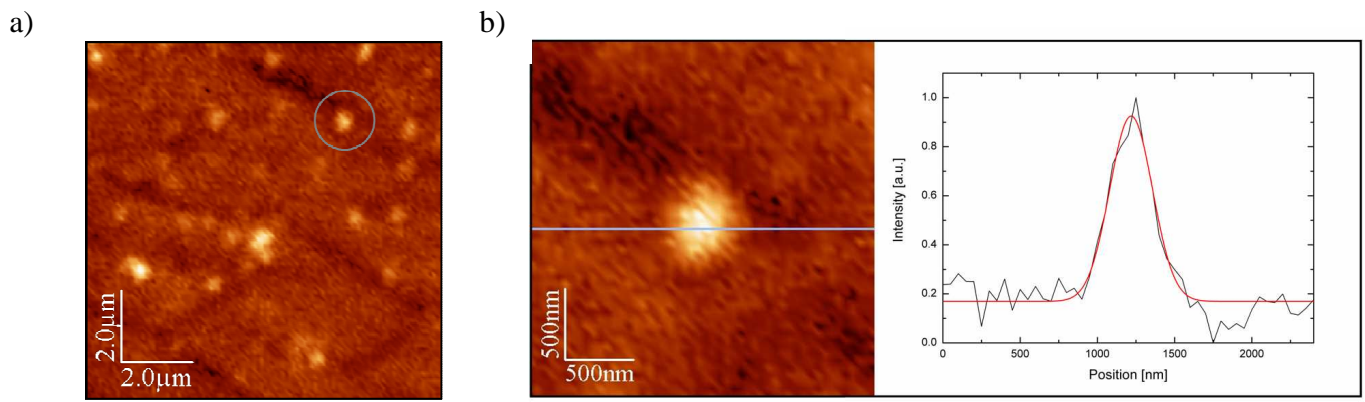

Figure 3. (a) Confocal image (10 µm x 10 µm) of a sample using vibronic excitation. Individual molecules can be seen as bright spots. (b) 2.5 µm x 2.5 µm scan over the molecule marked in (a). A Gaussian profile fitted to the cross section estimates the focus size to be as small as 330 nm.

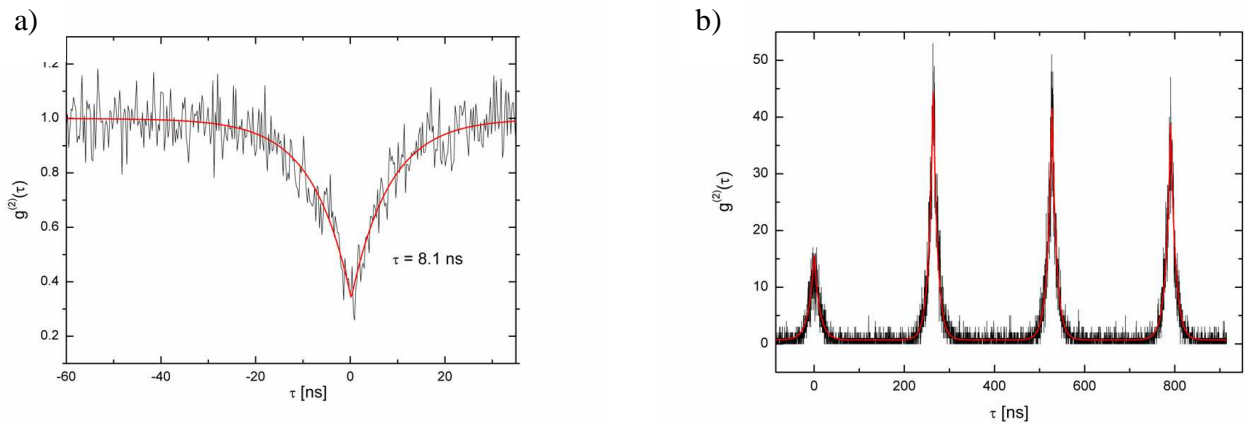

Figure 4. (a) Second order photon correlation measurement of the zero-phonon line of a molecule using cw excitation. (b) Triggered single photon emission by the same molecule using pulsed excitation. Strong antibunching is clearly visible in both cases.

a) 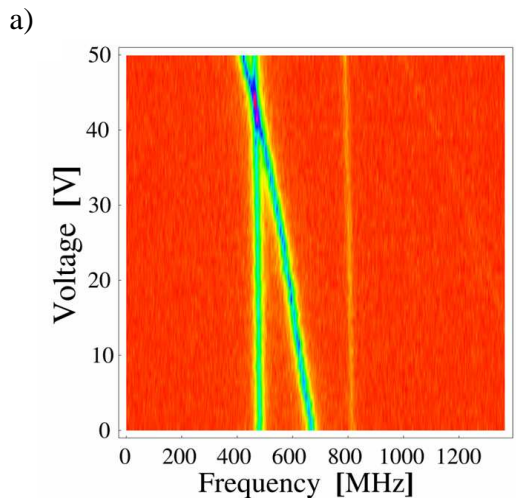 b) 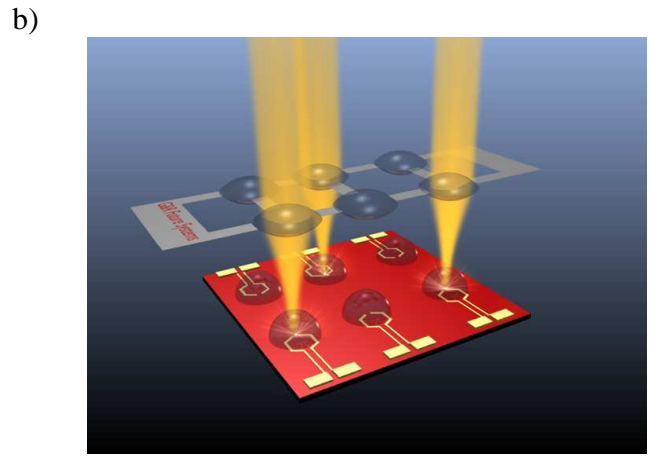

Figure 5.(a) Fluorescence excitation spectra of two molecules in different microscopes. An electric field applied to one of the molecules shifts its fluorescence to lower frequencies with increasing voltage. The two molecules become spectrally indistinguishable at a voltage of 42 V. (b) Schematics of multiple single-photon sources on a single chip where molecules are individually addressable by laser beams focused by aspheric lens/SIL combinations. Electrodes can be used to shift the emission wavelength of each single-photon source independently.